# Graph Theory Based Approach to Users Grouping and Downlink Scheduling in FDD Massive MIMO

Ali Maatouk*†, Salah Eddine Hajri*, Mohamad Assaad*, Hikmet Sari§†, and Serdar Sezginer†

*TCL Chair on 5G, Laboratoire des Signaux et Systèmes, CentraleSupélec, Gif-sur-Yvette, France
§NUPT, 66 Xinmofan Road, Gulou District, Nanjing, 210003 China
†Sequans Communications, 15 – 55 Boulevard Charles de Gaulle, 92700 Colombes, France

*Abstract*—Massive MIMO is considered as one of the key enablers of the next generation 5G networks. With a high number of antennas at the BS, both spectral and energy efficiencies can be improved. Unfortunately, the downlink channel estimation overhead scales linearly with the number of antenna. This does not create complications in Time Division Duplex (TDD) systems since the channel estimate of the uplink direction can be directly utilized for link adaptation in the downlink direction. However, this channel reciprocity is unfeasible for the Frequency Division Duplex (FDD) systems where different physical transmission channels are existent for the uplink and downlink. In the aim of reducing the amount of Channel State Information (CSI) feedback for FDD systems, the promising method of two stage beamforming transmission was introduced. The performance of this transmission scheme is however highly influenced by the users grouping and selection mechanisms. In this paper, we first introduce a new similarity measure coupled with a novel clustering technique to achieve the appropriate users partitioning. We also use graph theory to develop a low complexity groups scheduling scheme that outperforms currently existing methods in both sum-rate and throughput fairness. This performance gain is demonstrated through computer simulations.

## I. INTRODUCTION

MOBILE traffic demand has never been as high as it is today due to proliferation of smart-phones, video streaming, and other data-hungry applications. The next generation mobile networks should, therefore, cope with the necessity of higher throughput. One of the promising technologies to enable this higher throughput is Massive MIMO [1]. Compared to the currently deployed multi-user MIMO systems, massive MIMO incorporates a much larger number of antennas. This technology was shown to provide better performance in terms of energy efficiency and overall capacity [2] which made massive MIMO a hot research topic and a key component of future standards.

However, the high number of antennas complicates the channel estimation and feedback. The downlink channel estimation overhead burden comes from the fact that it scales linearly with the number of antennas [3]. This is alleviated in TDD systems by exploiting the channel reciprocity since the channel estimate of the uplink direction can be directly utilized for the downlink direction [3]. This is not possible for FDD systems that still represent the vast majority of currently deployed cellular networks.

To deal with this difficulty, the authors in [4] proposed Joint-Spatial-and-Division-Multiplexing (**JSDM**), an approach to multiuser MIMO downlink that is considered one of the most promising candidates for FDD massive MIMO. It works by partitioning users with the same second order downlink channel statistics into groups and splitting the downlink beamforming into two stages: an outer precoder, that depends on the channel statistics, and an inner precoder that depends on the instantaneous *effective* channel realizations. The role of the precoders being to suppress inter-group and intra-group interference respectively. The dimensions of the *effective* channel is significantly less than the number of antennas, thanks to the outer precoder projection. Even with this reduction in CSIT feedback, the authors in [4] showed that JSDM achieves the same sum capacity of the corresponding MU-MIMO downlink channel if the eigenspaces of groups are mutually orthogonal, a condition they called "tall unitary".

In realistic scenarios, users might have similar but not necessarily identical second order downlink channel statistics. This dictates the incorporation of a clustering algorithm to partition users into groups with *sufficiently* similar covariance eigenspaces. On top of that, with a high number of users uniformly distributed across the cell, the eigenspaces of the groups are far from meeting the tall unitary condition and a reduction of the number of simultaneously served groups is required. These issues inspired the work in [5] where $K$-means clustering algorithm was adopted and a greedy sum-rate maximization scheduling algorithm was proposed.

Due to the greediness nature of the scheduling scheme proposed in [5] and to simplify users grouping, recent work [6] adopted a hierarchical clustering algorithm which mixes both target number of clusters and chordal distance threshold to reach an appropriate users clustering. Average Signal-to-Leakage-plus-Noise-Ratio (SLNR) based scheduling approach was also proposed and was shown to outperform in terms of sum-rate all the previous methods in the literature [6].

In our paper, we deal with the issues that are still present in the previous approaches. First, the target number of clusters is not known beforehand and choosing an arbitrary number of clusters can have severe impact on the performance of JSDM. On top of that, appropriate chordal distance thresholds are hard to predict in the clustering process. This inspired us to propose a novel similarity measure along with a new clustering scheme where the number of clusters is not required to be known. Taking into account the lack of orthogonality between the eigenspaces of groups, adopting a scheduling scheme is of great importance to enhance the overall performance of JSDM. However, groups scheduling that aims to improve the average SLNR as in [6] does not necessarily translate into a higher sum-rate. Also, fairness is an issue that was not previously addressed. For instance, when adopting the approach in [6], some groups suffer from starvation. We therefore develop, by using graph theory tools, a low complexity scheduling scheme that outperforms all currently proposed methods in both sum-rate and throughput fairness.

The paper is organized as follows: Section II describes the system model. Section III presents the newly proposed metric and clustering method. Section IV includes the outer precoder design and the development of our scheduling scheme. Section V provides numerical results that demonstrate the performance of our method while Section VI concludes the paper.

## II. SYSTEM MODEL

We consider a single cell downlink multiuser MIMO system with $N_t$ antennas at the base station and $K$ single-antenna users. Let $\boldsymbol{y} \in \mathbb{C}^{K \times 1}$ be the received signal by the users:

$$\boldsymbol{y} = \boldsymbol{H}^H \boldsymbol{x} + \boldsymbol{z} \quad (1)$$

where $\boldsymbol{x} \in \mathbb{C}^{N_t \times 1}$ is the transmitted signal vector, $\boldsymbol{z} \in \mathbb{C}^{K \times 1}$ denotes the additive white Gaussian noise vector and $\boldsymbol{H} \in \mathbb{C}^{N_t \times K}$ is the channel matrix. The transmitted signal vector is actually a precoded version of the data vector $\boldsymbol{x} = \boldsymbol{V}\boldsymbol{d}$ where $\boldsymbol{V} \in \mathbb{C}^{N_t \times S}$ is the precoder and $\boldsymbol{d} \in \mathbb{C}^{S \times 1}$ is the data vector. The dimension S is equal to the number of total independent streams and is upper bounded by $min\{N_t, K\}$[4]. For the sake of simplicity, we adopt the approach of [4] with equal power allocation i.e. $\mathbb{E}(\boldsymbol{d}\boldsymbol{d}^H) = \frac{P}{S}\boldsymbol{I}_S$ where $P$ is the total downlink power budget. We suppose that $\boldsymbol{z} \sim \mathcal{CN}(0, \boldsymbol{I}_K)$. Assuming no line-of-sight propagation, we have $\boldsymbol{h}_k \sim \mathcal{CN}(0, \boldsymbol{R}_k)$ where $\boldsymbol{R}_k$ is a positive semi-definite covariance matrix. Let the eigenvalue decomposition (**EVD**) of $\boldsymbol{R}_k$ be the following:

$$\boldsymbol{R}_k = \boldsymbol{U}_k \boldsymbol{\Lambda}_k \boldsymbol{U}_k^H \quad (2)$$

where $\boldsymbol{\Lambda}_k$ is an $r_k \times r_k$ diagonal eigenvalues matrix with $r_k$ being the rank of $\boldsymbol{R}_k$ and $\boldsymbol{U}_k \in \mathbb{C}^{N_t \times r_k}$ being the set of eigenvectors corresponding to the non-zero eigenvalues. We adopt the two-stage precoders approach proposed in [4] where $\boldsymbol{V} = \boldsymbol{B}\boldsymbol{P}$. Based on the similarity of their channel covariance, users are partitioned into G groups with each containing $K_g$ users such as $K = \sum_{g=1}^{G} K_g$. The outerprecoder $\boldsymbol{B}$, of dimensions $N_t \times b$, is designed in a way to minimize the inter-group interference based on the channel statistics, which is supposed to be known at the base station as adopted in [4][1]. The inner precoder $\boldsymbol{P}$, of dimensions $b \times S$, depends on the instantaneous channel realizations and is intended to suppress intra-group interference. By taking into account the partitioning of users into $G$ groups, we have the following: $\boldsymbol{H}_g = [\boldsymbol{h}_{g_1}, \ldots, \boldsymbol{h}_{g_{K_g}}]$, $\boldsymbol{H} = [\boldsymbol{H}_1, \ldots, \boldsymbol{H}_G]$, $\boldsymbol{B} = [\boldsymbol{B}_1, \ldots \boldsymbol{B}_G]$, $\boldsymbol{P} = diag\{\boldsymbol{P}_1, \ldots, \boldsymbol{P}_G\}$ and we define the effective channel $\widetilde{\boldsymbol{H}} = \boldsymbol{B}^H \boldsymbol{H}$. It is straightforward that the effective channel is of dimension $b \times K$ with $b = \sum_{g=1}^{G} b_g$ and $b_g \ll N_t$. This drastically reduces the amount of CSI feedback in the case when each user $g_k$ feedback his *effective* channel $\widetilde{\boldsymbol{h}}_{g_k} \in \mathbb{C}^{b_g \times 1}$ rather than $\boldsymbol{h}_{g_k} \in \mathbb{C}^{N_t \times 1}$. We will refer to this approach as *per-group processing* (**PGP**).

[1]The channel statistics vary at a much slower rate than the channel coherence time and therefore can be assumed to be locally stationary and easily tracked by methods cited in [4]

The received signal by group g can be therefore written as:

$$\boldsymbol{y}_g = \boldsymbol{H}_g^H \boldsymbol{B}_g \boldsymbol{P}_g \boldsymbol{d}_g + \sum_{g' \neq g} \boldsymbol{H}_g^H \boldsymbol{B}_{g'} \boldsymbol{P}_{g'} \boldsymbol{d}_{g'} + \boldsymbol{z}_g \quad (3)$$

where $\boldsymbol{d}_g \in \mathbb{C}^{S_g \times 1}$ with $S_g$ being the number of independent streams intended for group $g$. By adopting the PGP approach and assuming perfect *effective* CSI at the BS, a zeroforcing (ZF) inner precoder can be calculated in the following manner:

$$\boldsymbol{P}_g = \zeta_g \widetilde{\boldsymbol{H}}_g (\widetilde{\boldsymbol{H}}_g^H \widetilde{\boldsymbol{H}}_g)^{-1} \in \mathbb{C}^{b_g \times S_g} \quad (4)$$

with $\zeta_g$ being a normalization factor to ensure that the power budget constraint is satisfied:

$$\zeta_g^2 = \frac{S_g}{tr(\boldsymbol{B}_g \widetilde{\boldsymbol{H}}_g \left(\widetilde{\boldsymbol{H}}_g^H \widetilde{\boldsymbol{H}}_g\right)^{-2} \widetilde{\boldsymbol{H}}_g^H \boldsymbol{B}_g^H)} \quad (5)$$

## III. CORRELATION CLUSTERING

In this section, we deal with the fact that users might have *similar* but not necessarily identical covariance matrices and appropriate grouping of users is essential for JSDM. The research papers that investigated this clustering problem presented two approaches: $K$-means clustering [5] and a hierarchical clustering [6]. Both of these approaches used the chordal distance as a metric. The downside of such a metric is the fact that prediction of any threshold involved in the clustering algorithm is a difficult task. Motivated by this, we propose in the following subsection a new similarity measure suitable for our problem.

### A. Similarity Measure

Herdin et al. [7] introduced a novel metric named Correlation Matrix Distance (**CMD**). It was used to track the changes of spatial structures of the channel in non-stationary MIMO. The use of this metric has been extended to many different research work. For example, the authors in [8] used it in the context of Grassmannian subspace packing. The same metric was also adopted by the authors in [9] to study the effect of subspace alignment in multi-user MIMO. The fact that CMD is normalized makes it more sensitive to differences in the correlation structure and present an opportunity in terms of threshold design. In the previous literature [5][6], the similarity between user 1 and user 2 was solely taken based on their covariance's eigenstructures $(\boldsymbol{U}_1 \boldsymbol{U}_1^H, \boldsymbol{U}_2 \boldsymbol{U}_2^H)$ without taking into account the energy of the modes. In our case, we will be applying our similarity measure on the whole covariance matrices $(\boldsymbol{R}_1, \boldsymbol{R}_2)$. The motivation behind this is that differences in the eigenstructures of weak modes should contribute less than the one's of strong modes. Based on CMD, we can define the new similarity measure as follows:

$$d_s(\boldsymbol{R}_1, \boldsymbol{R}_2) = 1 - CMD(\boldsymbol{R}_1, \boldsymbol{R}_2) = \frac{Tr(\boldsymbol{R}_1^H \boldsymbol{R}_2)}{||\boldsymbol{R}_1||_F . ||\boldsymbol{R}_2||_F} \quad (6)$$

This similarity measure is lowerbounded by 0 and upperbounded by 1. A value of 0 corresponds to the case where $\boldsymbol{R}_1$ and $\boldsymbol{R}_2$ are orthogonal while a value of 1 takes place when $\boldsymbol{R}_1$ and $\boldsymbol{R}_2$ are collinear. This proposed measure can be regarded as an extension of the cosine similarity of vectors (which is a widely used metric in clustering algorithms see, e.g., [10]) to matrices and therefore can now be considered as what we will call *Degree of OverLap* (**DOL**) between the two spaces. To our knowledge, this is the first time it has been used in the context of users clustering for FDD massive MIMO.

*B. Clustering Algorithm*

Unlike the previously proposed approaches, we seek to use a clustering algorithm without passing the target number of clusters as a parameter. To do so, we take advantage of the ease of threshold design presented by our proposed similarity metric. An interesting way to do so is by choosing $DOL_{th}$ high enough such as if $d_s(\boldsymbol{R}_k, \boldsymbol{R}_{k'}) \geq DOL_{th}$ then users $k$ and $k'$ can be thought to be laying in the same correlation space. Unlike other similarity metrics, this threshold is easily determined. One can simply say if the degree of overlap between the two spaces is above 0.95 then consider them as highly similar and are preferred to be assigned to the same cluster. Based on this, we can construct what we will call a *complete advice graph* $G_c = (V_c, E_c)$ where each vertex represents a user and an edge would have a $\langle +1 \rangle$ label to signal that these two users are preferred to be in the same cluster while $\langle -1 \rangle$ label refers to the opposite case.

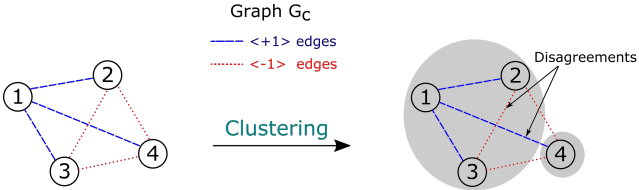

Fig. 1: Clustering Algorithm

Our goal now is therefore to produce a partition of the graph's vertices in a way that agrees as much as possible with the edge labels. To do so, we propose a cost function $J$ as the total disagreements of our resulting partitioned graph. The total disagreements cost is defined as the overall negative weights inside a cluster added to the positive weights between clusters. Our partitioning problem can be hence formulated as follows:

$$\begin{aligned}
\text{minimize} \quad & J = \sum_{(u,v) \in E_c^+} x_{uv} + \sum_{(u,v) \in E_c^-} (1 - x_{uv}) \\
\text{subject to} \quad & x_{uv} + x_{vw} \geq x_{uw} \quad \forall u, v, w \in V_c \\
& x_{uv} = x_{vu} \quad \forall u, v \in V_c \\
& x_{uv} = \begin{cases} 0 & \text{if } u \text{ and } v \text{ are in the same cluster} \\ 1 & \text{Otherwise} \end{cases}
\end{aligned} \quad (7)$$

The constraints take into account the symmetry of $x_{uv}$ and the triangular inequality[2] satisfied by these variables. What makes this clustering formulation interesting is that there is no need to specify the target number of clusters. Instead, the resulting optimal number of clusters could be any value from 1 to K depending on what fits our graph the most. Our problem in (7) turns out to be the same as the one studied in [11][12]. In general, solving (7) and finding the optimal clustering is NP-hard, as proven by the authors of [11] using a reduction from Exact Cover by 3-Sets (X3C) which is one of Karp's 21 *NP-complete* problems. One way to deal with this hardness is to turn the problem into a Linear Program **LP** by relaxing the binary condition and replacing it by $x_{uv} \in [0, 1] \ \forall u, v \in V_c$. The LP is then solved in polynomial time by any desired standard LP solvers followed by appropriate rounding of the fractional values. The question that remains is how to round the fractional values? The literature is rich with rounding techniques that achieves a good approximation ratio compared to the optimal solution. The most recent work in [12] proposed a new randomized technique based on *Pivoting*. Pivoting works by treating the fractional solution of (7) as a probability to put the two vertices in different clusters. The algorithm that was proposed in [12] is to apply the following function on the solution of (7) before proceeding to the pivoting phase:

$$f^+(x_{uv}) = \begin{cases} 0 & \text{if } x < a \\ (\frac{x_{uv}-a}{b-a})^2 & \text{if } x \in [a, b] \\ 1 & \text{if } x \geq b \end{cases} \quad f^-(x_{uv}) = x_{uv} \quad (8)$$

where $\langle + \rangle$ and $\langle - \rangle$ refer to $(u, v) \in E_c^+$ and $(u, v) \in E_c^-$ respectively. This rounding technique is guaranteed to achieve an expected (2.06-$\epsilon$)-approximation for $a = 0.19$, $b = 0.5095$ and a constant $\epsilon$ such as $0 < \epsilon < 0.01$. A derandomized version of the algorithm was also considered but we omit it for the sake of space and we refer the readers to [12]. Overall, the clustering algorithm can be summarized in Algorithm 1.

---

**Algorithm 1** Clustering Algorithm

1: **Init.** Compute the similarity matrix $\boldsymbol{S}_{ij} = d_s(\boldsymbol{R}_i, \boldsymbol{R}_j)$
2: **if** $s_{ij} > DOL_{th}$ **then** $s_{ij} = +1$
3: **else** $s_{ij} = -1$
4: **end if**
5: Solve (7) to get $x_{uv}$ then apply (8) to get $p_{uv} = f(x_{uv})$
6: **procedure** PIVOTING
7:     Let $V_0 = V_c$ the set of all vertices, let $t = 0$
8:     **while** $V_t \neq \emptyset$ **do**
9:         Pick a pivot $w_t \in V_t$ randomly and let $S_t = w_t$
10:         $\forall \, u \in V_t$, add $u$ to $S_t$ with probability $1 - p_{wu}$
11:         let $V_{t+1} = V_t \setminus S_t$, let $t = t + 1$
12:     **end while**
13: **end procedure**
14: Output the clusters $S_0, \ldots, S_{Final}$

---

[2]The triangular inequality ensures that if $(u, v)$ are assigned to the same cluster and $(v, w)$ are assigned to the same cluster, then definitely $(u, w)$ lay in the same cluster as well

## IV. Downlink Scheduling

After grouping users with similar second order channel statistics, we can now deal with the orthogonality aspect of JSDM. In realistic scenarios, groups do not lay in mutual orthogonal channel covariance spaces and inter-group interference can therefore limit the overall performance. One can seek to reduce this interference by applying appropriate outer precoding techniques but it is insufficient as will be proven shortly. Therefore, adopting a scheduling scheme is of paramount importance for the overall performance of JSDM.

### A. Problem Formulation

We define the centroid that would be taken as a representative of each group's equivalent covariance:

$$\boldsymbol{R}_g = \frac{1}{K_g} \sum_{k=1}^{K_g} \boldsymbol{R}_{g_k} \stackrel{\text{EVD}}{=} \boldsymbol{U}_g \boldsymbol{\Lambda}_g \boldsymbol{U}_g^H \text{ with } \boldsymbol{U}_g \in \mathbb{C}^{N_t \times r_g} \quad (9)$$

where $r_g$ is the rank of $\boldsymbol{R}_g$. This is where the clustering algorithm's effect is highlighted. Due to our clustering algorithm and with a high $DOL_{th}$, we know that the covariances of users in each group are really similar which means the centroid is a good representative of each cluster. However this is not necesarrily true in the case of other clustering algorithms with *pre-determined target number of clusters* as in [5][6] since it is hard to predict beforehand the right number of clusters for which each group's representative is a good one.

As proposed in [4] and adopted in [6], one can seek to eliminate inter-group interference by simply building up the inter-group interference matrix for group $g$ and projecting the intended signal space on the interference matrix's orthogonal space, a method that was given the name *Approximate Block Diagonalization*. The interference matrix seen by group $g$ is $\boldsymbol{\Xi}_g = [\boldsymbol{U}_1^*, \ldots, \boldsymbol{U}_{g-1}^*, \boldsymbol{U}_{g+1}^*, \ldots, \boldsymbol{U}_G^*]$ which is based on the eigenspace of other *active* groups with $\boldsymbol{U}_{g'}^* \in \mathbb{C}^{N_t \times r_{g'}^*}$ where $r_{g'}^*$ is a design parameter which denotes the number of the channel's strongest modes[3] taken into account. Let $[\boldsymbol{E}_g^{(1)}, \boldsymbol{E}_g^{(0)}]$ denotes the set of left eigenvectors of $\boldsymbol{\Xi}_g$ where $\boldsymbol{E}_g^{(0)}$ of dimensions $N_t \times (N_t - \sum_{g' \neq g} r_{g'}^*)$ form a unitary basis for $Span^\perp(\boldsymbol{U}_{g'}^* : g' \neq g)$. The projected channel covariance matrix is given by:

$$\widehat{\boldsymbol{R}}_g = (\boldsymbol{E}_g^{(0)})^H \boldsymbol{U}_g \boldsymbol{\Lambda}_g \boldsymbol{U}_g^H \boldsymbol{E}_g^{(0)} \stackrel{\text{EVD}}{=} \boldsymbol{G}_g \boldsymbol{\Phi}_g \boldsymbol{G}_g^H \quad (10)$$

The next step would be to match the $b_g$ strongest eigenmodes of our projected channel. Let $\boldsymbol{G}_g = [\boldsymbol{G}_g^{(1)}, \boldsymbol{G}_g^{(0)}]$ where $\boldsymbol{G}_g^{(1)}$ is of dimension $(N_t - \sum_{g' \neq g} r_{g'}^*) \times b_g$. Overall, the outer precoder is $\boldsymbol{B}_g = \boldsymbol{E}_g^{(0)} \boldsymbol{G}_g^{(1)}$. One would say that we can simply put $r_{g'}^* = r_{g'}$ i.e. includes all modes and we will lay in an interference free scenario. But by construction, we have that the effective channel dimension $b_g \leq rank(\widehat{\boldsymbol{R}}_g) = min(r_g, N_t - \sum_{g' \neq g} r_{g'}^*)$, which means including more modes would shrink our dimensionality and leads to a dimensional bottleneck. Keeping in mind that $S_g \leq b_g$, the dimensionality bottleneck is a serious matter since we have a certain number of independent streams that we are obligated to send out for each group. Since not necessarily all modes are included ($r_{g'}^* \leq r_g$), inter group interference would still be inevitable and we therefore include downlink scheduling to enhance the performance of the system. To formulate our scheduling problem, we have to adopt a certain network utility function.

The first building block of any network utility is the expression of the rate $R_{g_k}$ achieved by each user $g_k$ in the network. The expression of the rate can be concluded from the SINR expression by applying $R_{g_k}(SINR) = log_2(1 + SINR_{g_k})$[4].

*Large scale analysis:* We focus in our paper on the case of multi-user *massive* MIMO where the number of antenna and users $N_t, K \longrightarrow +\infty$. In this case, Random Matrix Theory (**RMT**) tools come in handy. The authors of JSDM [4] made use of the work in [13] to propose a deterministic equivalent for the $SINR$ expression in JSDM. Motivated by the fact that this deterministic equivalence was shown to be accurate for realistic values of $(N_t, K)$ [4][13], we take it as a basis of our analysis. Details concerning these equations can be found in [4] which for our case reduce to:

$$SINR_{g,k} \stackrel{N_t, K \to +\infty}{\longrightarrow} \frac{\frac{P}{S} x_g \bar{\zeta}_g^2}{\sum_{g' \neq g} \frac{P}{S} x_{g'} \bar{\zeta}_{g'}^2 \overline{\Upsilon}_{g,g'} + 1} \quad (11)$$

where $x_g$ is a binary variable that denotes if group $g$ is scheduled. $\bar{\zeta}_g^2 = \overline{m}_g b_g$, $\overline{\Upsilon}_{g,g'}$ and $\overline{m}_g$ are the results of fixed point equations with $\overline{\boldsymbol{R}}_g = \boldsymbol{B}_g^H \boldsymbol{R}_g \boldsymbol{B}_g$:

$$\overline{m}_g = \frac{1}{b_g} tr(\overline{\boldsymbol{R}}_g \boldsymbol{T}_g) \quad (12)$$

$$\boldsymbol{T}_g = \left( \frac{S_g}{b_g} \frac{\overline{\boldsymbol{R}}_g}{\overline{m}_g} + \boldsymbol{I}_{b_g} \right)^{-1} \quad (13)$$

$$\overline{\Upsilon}_{g,g'} = \frac{S_{g'}}{b_{g'}} \frac{n_{g',g}}{(\overline{m}_{g'})^2} \quad (14)$$

$$n_{g',g} = \frac{\frac{1}{b_{g'}} tr(\overline{\boldsymbol{R}}_{g'} \boldsymbol{T}_{g'} \boldsymbol{B}_{g'}^H \boldsymbol{R}_g \boldsymbol{B}_{g'} \boldsymbol{T}_{g'})}{1 - \frac{\frac{S_{g'}}{b_{g'}} tr(\overline{\boldsymbol{R}}_{g'} \boldsymbol{T}_{g'} \overline{\boldsymbol{R}}_{g'} \boldsymbol{T}_{g'})}{b_{g'} (\overline{m}_{g'})^2}} \quad (15)$$

What makes those expressions interesting is the fact that the effect of small-scale fading is averaged out as predicted in [2]. This is quite convenient since the second order statistics change at a much smaller rate than the channel coherence time as previously pointed out in Section II.

With the rate expression approximation dealt with, the goal now is to schedule these clusters in a way to get the highest utility, while preserving fairness between clusters and ensuring a certain *quality of link* for users in each group. Motivated by facilitating the design of the *quality of link* tolerance, we consider as a criterion the Signal-to-Interference Ratio $SIR_{g_k}$ experienced by *scheduled* users which is a widely used criterion in power control for wireless cellular networks [14].

$$SIR_{g_k} = \frac{\bar{\zeta}_g^2}{\sum_{g' \neq g} x_{g'} \bar{\zeta}_{g'}^2 \overline{\Upsilon}_{g,g'}} \quad (16)$$

---
[3]This is the reason why the energy of the modes was included in our clustering design previously since the strongest modes are taken into account first in the outer precoder design

The $SIR_{g_k}$ is forced to be above a certain threshold beyond which the link to user $g_k$ is supposed to be good. Since the $SIR_{g_k}$ of user $g_k$ depends *solely* on the group index $g$, we will drop the sub-index $g_k$ and work with $SIR_g$ experienced by the group $g$. Putting it all together, and taking the weighted sum-rate as utility, we can formulate our scheduling problem:

$$\begin{aligned}\text{maximize} \quad & \sum_{g=1}^{G} x_g \sum_{k=1}^{K_g} w_{g_k} R_{g_k} \\ \text{subject to} \quad & SIR_g \geqslant \alpha_g \ g=1,\ldots,G \\ & x_g \in \{0,1\} \ g=1,\ldots,G\end{aligned} \quad (17)$$

The previous studies in this area have not considered fairness between users (See, e.g., [6]). Introducing this weight $w_{g_k}$ allows us to incorporate fairness in our scheduling scheme. A special case would be the stable policy max-weight scheduling [15] where $w_{g_k}$ is chosen to be the length of the queue $Q_{g_k}$.

### B. Scheduling Scheme

In order to solve our problem in (17), we propose a 2-steps scheme based on graph theory. The first step deals with both the $SIR$ constraint and a combinatorial difficulty faced in our problem. The second step aims to find the appropriate combination of groups to be able to solve (17). From the $SIR_g$ expression in (16), we can construct a weighted *directed* graph $G_L = (V, E)$ where V is the set of groups and in which the weight of the edge $e(g', g)$ corresponds to what we will call the *normalized interference* from group $g'$ to group $g$:

$$e(g', g) = \frac{\bar{\zeta}_{g'}^2 \overline{\Upsilon}_{g,g'}}{\bar{\zeta}_g^2} \quad (18)$$

*One thing to point out is that due to the fact that the outer precoder $B_g$ depends on the eigenspace of all active groups, the weight of the edges in this graph are not* constant *and depend on the activity of all groups as seen from the fixed point equations (12)-(15). This complicates our problem since in normal graphs, the weight of the edges is fixed and do not change as you manipulate the graph. To alleviate this issue and deal with the SIR constraint in (17), we first proceed to what we will call the "Elimination" phase. This phase also allows us to convert our problem into a vertex coloring problem, as will be detailed in the "Grouping" phase. An example of 4 groups scenario will be presented in successive figures to fully demonstrate the mechanisms of the scheduling scheme.*

*1) Elimination:* We can picture each vertex in the graph as a sink of interference that undergoes successive *iterations*. In the first iteration, all groups are considered to be active. The outerprecoder $B_g$ of each group is calculated as detailed in Section IV-A. The fixed points equations (12)-(15) are then solved and $e_1(g', g) \ \forall (g', g) \in V^2$ are calculated based on (18) where the sub-index "1" refers to the iteration number. For each vertex $g \in V$, the $SIR_g$ condition of (17) is tested. If it is violated, the edge $e_1(g', g)$ with the highest weight is eliminated. In other words, the group that interfere most with $g$ is chosen to be eliminated. At the next iteration, we have a new graph due to the edges removal from the previous iteration and therefore the outer precoders are to be recalculated. This time however, the outer precoder $B_g$ of each vertex $g \in V$ is calculated based on the eigenspace of neighboring[4] vertices only. We repeat the same procedures of the first step: the fixed points equations are solved again (12)-(15) and the weight of the edges of *neighboring* vertices only are recalculated using (18). The stopping criteria would be if an iteration resulted in no new deleted edges. In other words, if simultaneous scheduling of *neighboring* vertices in the resulting graph will not vioalate the $SIR$ condition of each of them. An example of the above procedure is given in Fig. 3, the first iteration resulted in four deleted edges. The edges of neighboring vertices are then updated for the second iteration. The second iteration did not result in any deleted edges and the algorithm finishes. Once the algorithm finishes, we turn our graph into an *undirected* graph $G_u = (V, E_u)$ by simultaneous agreements from both sides i.e. if $e(g, g')$ and $e(g', g)$ are both not eliminated in $G_L$ then an edge $e_u(g, g') = 1$ exist in $G_u$ and $e_u(g, g') = 0$ otherwise. In our new *undirected* graph $G_u$, an edge exist between two vertices if scheduling them together will not violate their respective $SIR$ conditions. We can now tackle another aspect of our problem: *Which groups of those that are allowed to transmit simultaneously should we schedule in order to maximize our utility?*

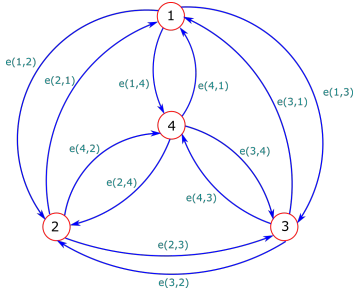

Fig. 2: Weighted Directed Graph $G_L$

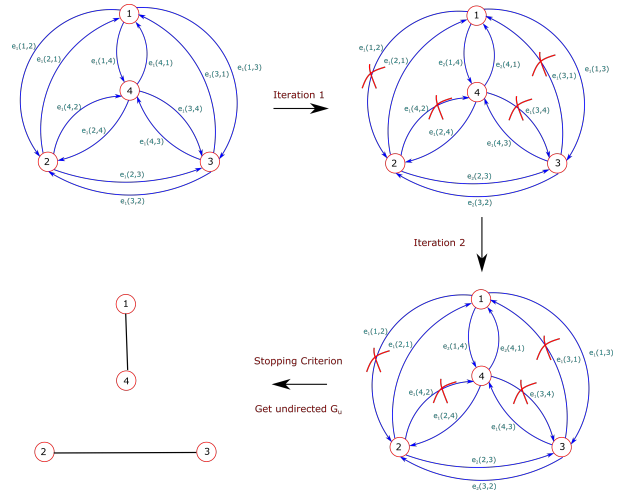

Fig. 3: Elimination Process

---

[4] A vertex $g$ is a neighbor of $g'$ at iteration $t+1$ if both directed edges $e_t(g, g')$ and $e_t(g', g)$ at iteration $t$ were not eliminated

*2) Grouping:* After proceeding with the elimination step, our $SIR$ constraint can be replaced by making sure that two simultaneously scheduled groups should have an edge between them in $G_u$. Therefore, our problem in (17) is turned into:

$$\begin{aligned} \text{maximize} \quad & \sum_{g=1}^{G} x_g \sum_{k=1}^{K_g} w_{g_k} R_{g_k} \\ \text{subject to} \quad & x_g + x_{g'} \leq 1 \ \forall (g,g') \notin E_u \\ & x_g \in \{0,1\} \ g=1,\ldots,G \end{aligned} \quad (19)$$

To solve this new problem, we recall that for a well chosen $\alpha_g \ \forall g$, an edge exist between two vertices in $G_u = (V, E_u)$ only if they barely interfere and hence scheduling them together would normally increase their sum utility. Our goal is therefore to find combinations of groups that are adjacent one to the other in $G_u$ while covering the whole vertex set $V$. For this purpose, we define a *clique* in an undirected graph as a subset of vertices such as every two distinct vertices in the clique are adjacent. We seek to find the *smallest number* of cliques that cover $V$, where we emphasize "smallest" to ensure that each clique have the largest number of groups possible inside. Essentially, we are trying to solve the *minimal clique vertex cover* problem. The *minimal clique vertex cover* problem is known to be equivalent to vertex coloring on the complement graph $\bar{G}_u$, a well known *NP-Complete* problem. Knowing that vertex coloring seeks to partition the set of vertices into the smallest number of independent sets, one can see the connection between the two problems since a subset of vertices is a clique in $G_u$ if and only if it is an independent set in $\bar{G}_u$. We will use a simple yet effective maximal independent set based vertex coloring algorithm that achieves a $O(\frac{n}{log(n)})$-approximation ratio [16] presented in Algorithm 2 and apply it on $\bar{G}_u$. After applying Algorithm 2, each group $g$ will be assigned a color $Col(g)$. Groups that are assigned the same color represent a subset of groups that are allowed to transmit simultaneously. We can therefore replace the edges constraint in (19) by ensuring that the color assignments are respected:

$$\begin{aligned} \text{maximize} \quad & \sum_{g=1}^{G} x_g \sum_{k=1}^{K_g} w_{g_k} R_{g_k} \\ \text{subject to} \quad & x_g + x_{g'} \leq 1 \ if \ Col(g) \neq Col(g') \\ & x_g \in \{0,1\} \ g=1,\ldots,G \end{aligned} \quad (20)$$

The problem in (20) is indeed simple to solve. One can simply form "*Schedules*", each made of groups that belong to the same color. At the start of each coherence time $T_c$, the schedule (i.e. the color) that leads to the largest utility is selected.

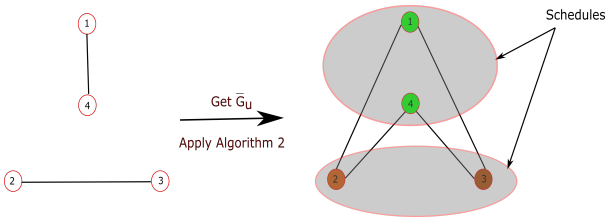

Fig. 4: Coloring Process

**Algorithm 2** Coloring Algorithm

1: **Initialization:** Let $S = \emptyset$ and index $i = 1$
2: **while** $S \neq V$ **do**
3:     Let $C_i = \emptyset$ the set of color $i$
4:     Let $R = V \setminus S$ the set of remaining vertices
5:     **while** $R \neq \emptyset$ **do**
6:         Pick a vertex $w \in R$ with lowest degree randomly and let $C_i = C_i \cup \{w\}$
7:         $R = R \setminus \{w\} \cup Neighbor(w)$
8:     **end while**
9:     $S = S \cup C_i$
10:     Output the color $C_i$ and $i \leftarrow i + 1$
11: **end while**

## V. NUMERICAL RESULTS

We consider a base station with a $120°$ sector centered around the x-axis consisting of a ULA with $N_t = 128$ antennas and serving $K = 80$ users arbitrarily distributed in the sector. For ease of correlation entries calculation, we adopt the one-ring model [4]. We consider a user terminal (UT) at an azimuth angle $\theta$ and angular spread $\Delta$. The correlation entry is then calculated for $1 \leqslant m,p \leqslant N_t$ using the following formula:

$$[\boldsymbol{R}]_{m,p} = \frac{1}{2\Delta} \int_{\theta-\Delta}^{\theta+\Delta} e^{j\boldsymbol{k}^T(\alpha)(\boldsymbol{u}_m - \boldsymbol{u}_p)} d\alpha \quad (21)$$

where $\boldsymbol{k}(\alpha) = \frac{-2\pi}{\lambda}(cos(\alpha), sin(\alpha))^T$ denotes the wave vector for a planar wave with angle of arrival $\alpha$, $\lambda$ is the wavelength and $\boldsymbol{u}_m, \boldsymbol{u}_p \in \mathbb{R}^2$ are the position vectors of the BS antennas in the 2D-coordinate system. For our scenario, we consider that all users have the same angular spread of $\Delta = 5°$. We set the clustering high threshold to $DOL_{th} = 0.95$. We suppose that $S_g = K_g$ and set $r_g^* = K_g$ as adopted in [6].

The goal of the simulations is to compare our method to the recently proposed SLNR based scheduling scheme [6]. We take the SLNR based scheduling scheme as a benchmark since it was shown to outperform all proposed methods in the litterature in terms of sum-rate [6]. We also expose the necessity of a scheduling policy by simulating JSDM without any scheduling just as in [4]. Due to the fact that both the SLNR based scheduling and our scheduling scheme require a threshold tolerance, we iterate over a wide range of thresholds and choose the one that led to the highest sum-rate as a representative of each methods for a fair comparison. To be able to compare our method to theirs, we create our *schedules* and switch between them in Symmetrical Round-Robin manner and take their average sum-rate as a representative while baring in mind that higher sum-rate can be achieved for our scheme by simply choosing the schedule with the highest sum-rate.

Fig. 5 shows how the JSDM method without scheduling performs poorly due to limitations in terms of inter-group interference which is a proof of the necessity of adopting a scheduling scheme. In addition, we can see how our method was able to outperform the SLNR based method over the whole $SNR$ range. This is because we work on the inter-ference itself to improve the sum-rate. On the other hand, increasing the average SLNR of a system as in [6] does not necessarily translate into a higher sum-rate.

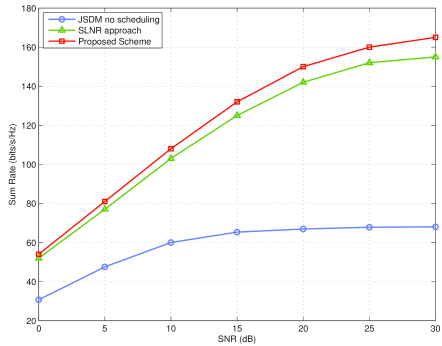

Fig. 5: Comparison of sum spectral efficiency vs. SNR

For the comparaison in terms of throughput fairness between the methods, the well known Jain's fairness index [17] was taken as a metric. It is defined as follows:

$$F(R_1, R_2, \ldots, R_K) = \frac{(\sum_{k=1}^{K} R_k)^2}{K \sum_{k=1}^{K} (R_k)^2} \qquad (22)$$

The values of this index range from $1/K$ (the case where only one user acquire the channel while the others end up starving) to 1 (the case where resources are shared equally between users). Fig. 6 shows that the SLNR based scheduling scored the worst fairness index due to the fact that after successive elimination of groups with low SLNR, the users inside these groups end up starving. The throughput fairness of JSDM with no scheduling is high but not perfect since users suffer different interference conditions and therefore asymmetric throughput. Our method scored almost perfect throughput fairness due to two reasons: the first being that by construction, $SIR$ of each group was chosen to be lower bounded by the same well chosen tolerance and the second being that symmetrical Round-Robin was adopted between schedules. *Overall, our proposed scheme was able to outperform the SLNR based method in sum-rate while providing a huge gain in terms of throughput fairness.*

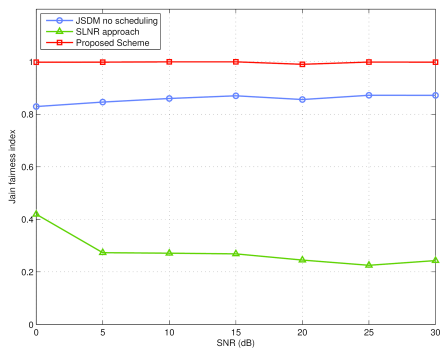

Fig. 6: Comparison of Jain's fairness index vs. SNR

## VI. CONCLUSION

In this paper, we tackled the problem of users clustering and scheduling in the promising technique of two-stage beamforming for the downlink of FDD massive MIMO. We introduced a new similarity metric coupled with a clustering method that are characterized by ease of design and good performance. We also developed using graph theory tools an efficient groups scheduling scheme that outperforms the current methods proposed in the literature in both sum-rate and throughput fairness as was shown in the simulations.